\documentclass[10pt]{article}
\usepackage[totalwidth=450pt, totalheight=590pt, centering]{geometry}

\usepackage{amsmath,amssymb,mathrsfs}
\usepackage{bbm} 

\usepackage[dvips]{graphicx}

\usepackage[dvips,bookmarks=false]{hyperref} 
\hypersetup{pdfstartview=FitH,pdfhighlight=/O,colorlinks=false}

\newcommand{\be}{\begin{equation}}
\newcommand{\ee}{\end{equation}}
\newcommand{\ben}{\begin{equation*}}
\newcommand{\een}{\end{equation*}}

\newcommand{\mc}[1]{\mathcal{#1}}

\newcommand{\mbb}[1]{\mathbb{#1}}

\newcommand{\p}{\partial}

\newcommand{\mdots}{,.\,.\,,}

\title{\Large{\textbf{Black Hole Entropy, Loop Gravity, and Polymer Physics}}}

\author{Eugenio Bianchi}

\author{Eugenio Bianchi\footnote{email: {\tt bianchi@cpt.univ-mrs.fr}}\\[.35em]
\normalsize{\emph{Centre de Physique Th\'eorique de Luminy{$\,$}\footnote{Unit\'e mixte de recherche (UMR 6207) du CNRS et des Universit\'es de Provence (Aix-Marseille I), de la M\'editerran\'ee (Aix-Marseille II) et du Sud (Toulon-Var); laboratoire affili\'e \`a la FRUMAM.} , case 907, F-13288 Marseille, France}}
}

\date{November 25, 2010}

\begin{document}

\maketitle

\begin{abstract}
Loop Gravity provides a microscopic derivation of Black Hole entropy. In this paper, I show that the microstates counted admit a semiclassical description in terms of shapes of a tessellated horizon. The counting of microstates and the computation of the entropy can be done via a mapping to an equivalent statistical mechanical problem: the counting of conformations of a closed polymer chain. This correspondence suggests a number of intriguing relations between the thermodynamics of Black Holes and the physics of polymers.
\end{abstract}


\section{Introduction}
Black Holes have thermodynamic properties \cite{Bekenstein:1973ur} (see \cite{Wald:1999vt} for a recent review). As is well known, the entropy $S$ of a Black Hole at equilibrium depends only on its mass $M$ and angular momentum $J$, and is proportional to the area $A_H$ of the horizon of the Black Hole\footnote{Throughout the paper, we will assume that the Black Hole is not charged and that there are no long range fields besides gravity. For a charged Black Hole, the horizon area and the entropy depend also on the charge. In the following, $G$ is Newton's constant, $\hbar$ is the reduced Planck constant, and $\kappa$ the Boltzmann constant. The speed of light $c$ is set to one.},
\begin{equation}
S(M,J)=\frac{\kappa}{4 G \hbar}\,A_H(M,J)\;.
\label{eq:S BH}
\end{equation}
This result due to Bekenstein and Hawking leads to a conceptual puzzle: what is the statistical mechanical origin of this entropy? what are the microstates responsible for heat exchanges of the Black Hole with its surroundings?

Within Loop Quantum Gravity (LQG), there is a large body of work regarding the derivation of the entropy of a Black Hole \cite{Corichi:2009wn}, starting with the first pioneering works of the late nineties \cite{Rovelli:1996dv}. The presence of a Black Hole is coded in an isolated-horizon boundary condition for the theory, as first discussed by Ashtekar, Baez, Corichi and Krasnov in \cite{Ashtekar:1997yu}. On the isolated horizon, a Chern-Simons theory with punctures is induced. The degrees of freedom counted are the states of this topological quantum field theory. For a large Black Hole, the counting amounts to the computation of the dimension of a certain $SU(2)$ intertwiner space, as recently discussed by Engle, Noui and Perez \cite{Engle:2009vc}. The entropy is given by the logarithm of the number of accessible states and results to be proportional to the area of the isolated horizon, thus reproducing the Bekenstein-Hawking formula (\ref{eq:S BH}).\\

The success of the LQG calculation raises a number of questions. What is the physical interpretation of the horizon degrees of freedom counted in LQG? Are they  present already at the classical level? are they purely quantum? In \cite{Rovelli:1996ti}, Rovelli first suggested that the degrees of freedom counted in LQG are in fact ``quantum shapes of the horizon''. The idea is discussed in more detail in a recent paper by Rovelli and Krasnov \cite{Krasnov:2009pd}. 

In this paper, I further develop this idea. I present a semiclassical description of the microstates counted in LQG: it is a description in terms of classical shapes of a \emph{tessellated horizon}. I show that the problem of counting the number of shapes can be mapped into an equivalent statistical mechanical problem: the counting of conformations of a closed polymer chain. This correspondence suggests a number of intriguing relations between the thermodynamics of Black Holes and the physics of polymers.\\

The presentation is largely independent from the full machinery of LQG. It relies on the recent development of the notions of \emph{coherent intertwiners} \cite{Livine:2007vk} and of the associated classical system \cite{Bianchi:2010gc,Hal}. The paper is organized as follows: in sections \ref{sec:shape} and \ref{sec:tessellated}, I review the idea that the entropy of a Black Hole is due to quantum fluctuations of its horizon geometry, and discuss the physical cut-off of horizon fluctuations provided by LQG; in \ref{sec:counting shapes}, \ref{sec:mapping} and \ref{sec:calculation}, I formulate the statistical mechanical problem, explain the mapping to a polymer chain, and present a simple calculation of the entropy.  Finally, I mention some possible further developments in sections \ref{sec:stretching} and \ref{sec:conclusions}.

\section{Black Hole entropy and the shape of the horizon}\label{sec:shape}
The idea that the microstates responsible for the entropy of a Black Hole correspond to fluctuations of the shape of its horizon goes back to Sorkin \cite{Sorkin:1983a} and to York \cite{York:1983zb}. The idea is the following. Let us focus on the non-rotating case $J=0$ for the moment. At equilibrium, the Black Hole is described by a Schwarzschild geometry. From a statistical mechanics point of view, this is to be understood as the \emph{macrostate}. While the macrostate is spherically symmetric, the microstates don't have to. For instance, in presence of a planet orbiting around a Black Hole -- because of tidal effects -- a small bulge is produced on the horizon. Similarly, in the presence of a gas at a finite temperature, the horizon will be thermally fluctuating. Even in absence of a thermal bath, the horizon will be fluctuating because of quantum effects. The key idea is that these horizon degrees of freedom are accessible from the exterior of the Black Hole, contribute to heat exchanges of the Black Hole with its surroundings, and provide a microscopic explanation of its thermodynamic entropy (see \cite{Jacobson:2005kr} for a recent discussion of this idea). 

The idea of horizon shapes resonates with another development started by Damour in the late seventies: the \emph{membrane paradigm} \cite{Damour:1982mg}. This is a description of the interaction of a Black Hole with the outside world in terms of a horizon boundary-condition with an fascinating physical interpretation. The boundary dynamics turns out to be the one proper of a (fictitious) physical membrane with mechanical properties and with a finite surface viscosity.

The fact that the entropy of a Black Hole can be accounted for considering the quantum fluctuations of the horizon was explored by Maggiore in the mid-nineties \cite{Maggiore:1994ww}. A perturbative Quantum-Field-Theory calculation of the fluctuations of the horizon/membrane leads in fact to an entropy proportional to the area of the horizon. However, the calculation requires an ultraviolet cut-off that shows up in the proportionality constant. When the cut-off is removed, the entropy diverges.

The idea explored in this paper is that LQG provides a very specific physical cut-off for the horizon-shape degrees of freedom. The physical cut-off is due to quantum geometry effects \cite{Rovelli:1994ge} proper of LQG, and makes the entropy finite.

\section{The LQG physical cut-off: a tessellated horizon}\label{sec:tessellated}
\begin{figure}%
\begin{center}
\includegraphics[width=.4\textwidth]{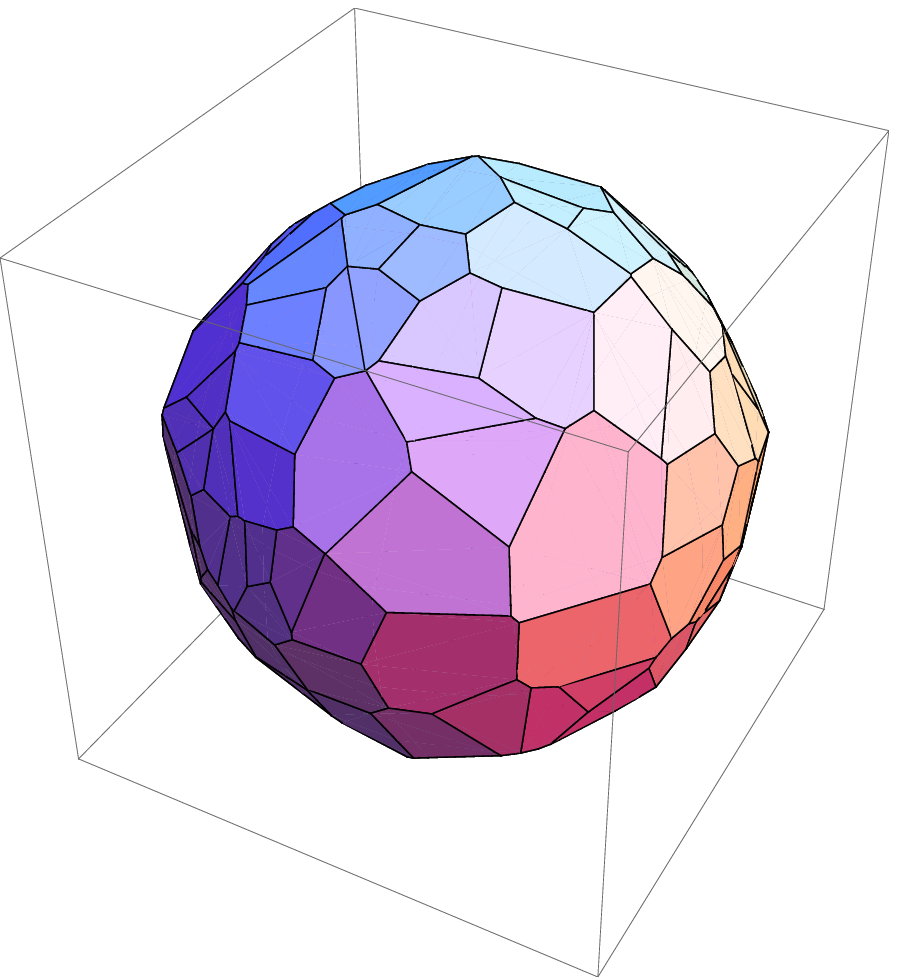}%
\hspace{3em}
\includegraphics[width=.45\textwidth]{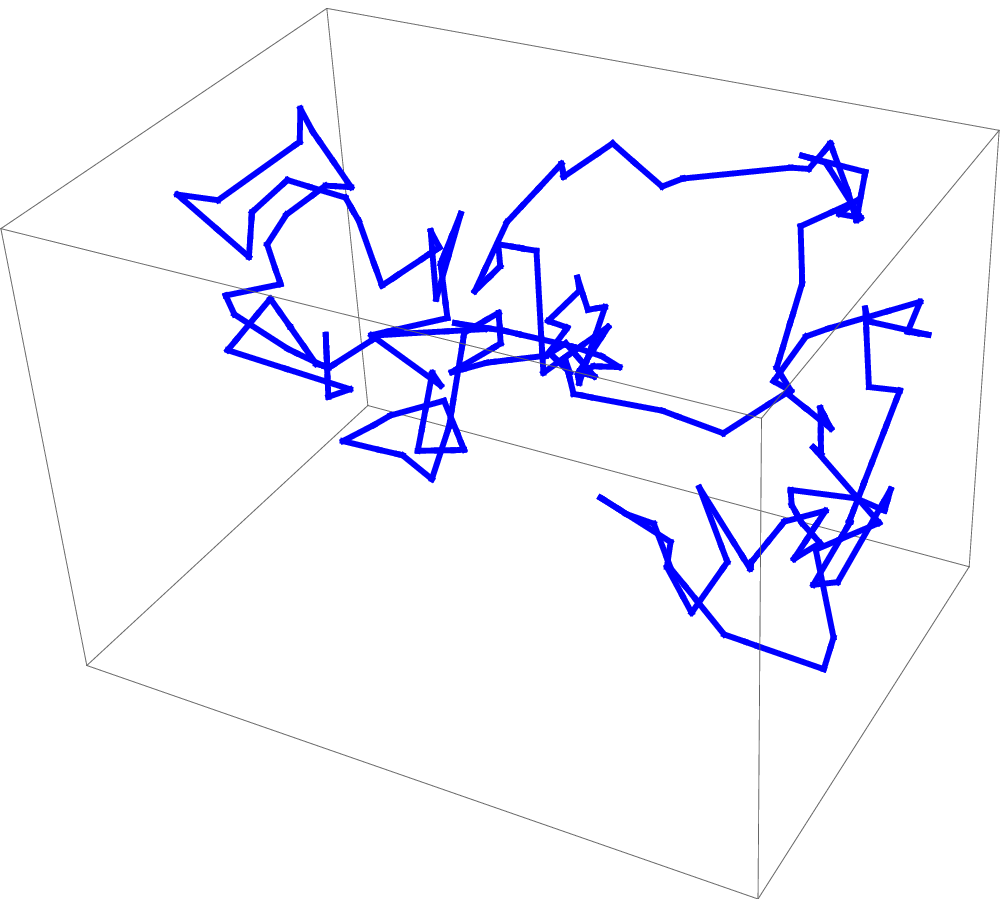}	
\end{center}
\caption{On the left: tessellated surface determined by the areas and normals of its facets; typical configuration belonging to the ensemble with fixed total area discussed in the text ($A_H\approx 100 \,L_P^2$). On the right: closed polymer chain corresponding to the tessellated surface.}%
\label{fig:tessellated}%
\end{figure}

Let us consider a set of $N$ positive real numbers $A_a\;$ ($a=1\mdots N$)  and $N$ unit-vectors $\vec{n}_a$ in $\mbb{R}^3$ that satisfy the \emph{closure} condition $\sum_a A_a\,\vec{n}_a\,=\,0$.
Thanks to a theorem due to Minkowski \cite{Minkowski}, we know that there exists a unique convex polyhedron in $\mbb{R}^3$ that has $N$ faces of area $A_a$ and normal $\vec{n}_a$ (see \cite{Bianchi:2010gc} for a recent discussion). The boundary is a surface with topology $S^2$ and endowed with a locally-flat geometry: it is a tessellated surface with $N$ facets. Its intrinsic and extrinsic geometry is completely determined by the data $(A_a,\;\vec{n}_a)$. In particular, the area of the facet $a$ is $A_a$, and the extrinsic curvature is coded in the angle between the normals $\vec{n}_a$ and $\vec{n}_b$ associated to two adjacent facets $a$ and $b$. See Fig. \ref{fig:tessellated}.

Recently, in a paper with Don\'a and Speziale \cite{Bianchi:2010gc}, we have shown that this tessellated surface arises as the classical limit of the quantum geometry proper of $SU(2)$ intertwiner space. The relation is briefly explained below.

Consider a system of $N$ particles of spin $j_a$, $(a=1\mdots N)$.
The Hilbert space of the system is the tensor product of the single particle Hilbert spaces, $\mc{H}^{(j_1)}\otimes \cdots \otimes \mc{H}^{(j_N)}$. An orthonormal basis is given by simultaneous eigenstates of the third component of the angular momentum operators $\vec{J}_a$, and of the Casimir operators $\vec{J}_a^2$. The system can be in a state $|s\rangle$ that is invariant under rotations. Rotationally invariant states satisfy the operator equation
\begin{equation}
\sum_{a=1,..,N}\vec{J}_a\;|s\rangle\;\,=\;0\;,
\label{eq:q closure}
\end{equation}
and form a subspace of the full Hilbert space. This subspace is called intertwiner space, $\mc{H}_0^{(j_1\mdots j_N)}$. In a recent paper \cite{Krasnov:2009pd}, Rovelli and Krasnov identify the quantum geometry of the horizon with states in this intertwiner space. The Casimir operator $\vec{J}_a^2$ measures the area of a facet of the quantum horizon. To be precise, each facet has an area that is quantized and given by
\begin{equation}
A_a=8\pi\,\gamma L_P^2\;\sqrt{j_a(j_a+1)}\;.
\label{eq:Aa}
\end{equation}
Here $j_a=\frac{1}{2},1,\frac{3}{2},\ldots$ is a $SU(2)$ representation associated to the facet. $L_P$ is a Planck length-scale and $\gamma$ is a positive dimensionless number, the Immirzi parameter. $L_P$ and $\gamma$ are to be understood as the two fundamental constants of LQG. The lowest non-vanishing value the area of a facet can have is the  Planck-scale gap $a_0=\frac{\sqrt{3}}{2}\,8\pi\gamma L_P^2$. It provides a physical ultraviolet cut-off for the horizon degrees of freedom. The quantum extrinsic curvature of the horizon is measured by the operators $\vec{J}_a\cdot \vec{J}_b$  (with $a\neq b$) and has again a discrete spectrum \cite{Major:1999mc}. Notice that these operators do not commute among themselves, and this fact leads to Heisenberg uncertainty relations for the quantum geometry of the horizon.\\

\noindent The relation between the two structures discussed above,
\begin{itemize}
	\item[(a)] the tessellated surface
	\item[(b)] the quantum geometry of intertwiners
\end{itemize} 
can be understood in two complementary ways. The first is to consider coherent states in the intertwiner space (b). Such states have been developed by Livine, Speziale, Freidel, Conrady and Krasnov in \cite{Livine:2007vk}, and are peaked on the geometry of a tessellated surface \cite{Bianchi:2010gc}. The semiclassical limit of the operator $\vec{J}_a$ is simply a classical vector in $\mbb{R}^3$ with norm $j_a$, and the operator equation (\ref{eq:q closure}) becomes the classical closure condition
\begin{equation}
\sum_{a=1,..,N}j_a\,\vec{n}_a\;=\;0\;,
\label{eq:j closure}
\end{equation}
thus reproducing the data that describe the system (a).
The second way to understand the relation between the two systems is to consider the quantization of the classical system (a). This has been done by Kapovich, Millson\footnote{The classical system considered in these papers is mathematically equivalent to (a), but the geometrical interpretation they describe differs from the one considered in this section.} \cite{Kapovich} and Charles \cite{Charles}, and shown to be the Hilbert space of intertwiners (b).\\

To summarize: the microstates of the horizon considered in LQG admit a semiclassical description. They consist of tessellated surfaces described by the following data: a set of $N$ vectors $j_a\vec{n}_a$ associated to each facet of the surface. Such data have to satisfy the closure condition (\ref{eq:j closure}) and the norms $j_a$ of the vectors have to be half-integers. The area of a facet is determined by the $j_a$s and can assume only a discrete series of values given by formula (\ref{eq:Aa}).

\section{The statistical-mechanical problem: counting shapes of a tessellated horizon}\label{sec:counting shapes}
Now we are ready to formulate our statistical-mechanical problem. Let us focus on an uncharged non-rotating Black Hole. The \emph{macrostate} of the system is a Schwarzschild geometry and is completely described by a single parameter: for instance the area $A_H$ of the horizon,
\begin{equation}
A_H=16\pi\, (GM)^2\;.
\label{eq:AH schw}
\end{equation}
The \emph{microstates} to be counted are shapes of the tessellated horizon described in terms of the data $A_a$ and $\vec{n}_a$. Only the shapes that are accessible have to be counted. The ensemble we consider here is the one of given average mass or, equivalently, given average horizon area $A_H$,
\begin{equation}
\textstyle\big\langle \sum_a A_a\big\rangle\;=\;A_H\;.
\label{eq:<A>}
\end{equation}
We call $W(A_H)$ the number of distinguishable microstates in the ensemble corresponding to horizon area $A_H$. The horizon entropy is then given by its natural logarithm,
\begin{equation}
S(A_H)=\kappa\,\log W(A_H)\;.
\label{eq:S(AH)}
\end{equation}
Notice that, while the macrostate has spherical symmetry, the microstates do not. This is a common feature in statistical mechanics: the microstates of the gas in this room are not translationally invariant.

\section{Mapping to an equivalent statistical-mechanical problem: conformations of a polymer chain}\label{sec:mapping}
What is the total number of configurations $W(A_H)$ ? At first sight, this counting problem looks extremely complicated. I understand that people in statistical mechanics have a standard strategy to address this kind of problems. The first question to be asked is if there is another system that is equivalent to the one considered here, and for which the counting has already been done.

It turns out that we are exactly in this situation: the equivalent statistical-mechanical problem here is the counting of conformations of a polymer. The mapping between the two systems  is such that the length of each monomer is given by the area of a facet of the tessellated horizon. 

The physics of polymers is well-studied (we refer the reader to the beautiful book by De Gennes \cite{DeGennes}). A basic result in polymer physics is that the entropy of a polymer is \emph{proportional to its length}, that is the sum of the lengths of its monomers. But this is also the sum of the areas of the facets of the tessellated horizon, that sum up to the area of the horizon of the Black Hole. Thus the Bekenstein-Hawking area law (\ref{eq:S BH}) is easily reproduced starting from the length-law proper of polymer physics. This is the main idea; in the rest of this section, I give a more precise and detailed description of the correspondence between shapes of the tessellated horizon and conformations of the polymer chain.  In next section, I present a simple derivation of the entropy.\\

\noindent The construction of the equivalent system is the following\footnote{The phase space of polygons studied by Kapovich and Millson in \cite{Kapovich} is directly related to this construction.}:
\begin{itemize}
	\item to a surface with $N$ facets, we associate a polymer chain consisting of $N$ monomers; 
	\item to a facet with normal $j_a \vec{n}_a$, we associate a monomer with direction $j_a \vec{n}_a$ in $\mbb{R}^3$; its norm $j_a$ is the length of the monomer;  	
	\item the fact that the horizon is a \emph{closed} tessellated surface corresponds to the requirement that the monomers belong to a closed polymer chain; their directions satisfy the condition (\ref{eq:j closure}).
\end{itemize}
Now some comments about the relation between the two systems, the Black Hole horizon and the closed polymer chain:
\begin{itemize}
	\item[-] the entropy of a polymer is finite because of the physical cut-off provided by the monomer chemistry; analogously in LQG the entropy of a Black Hole is finite because of the Planck scale cut-off provided by the quantization of the area;
	\item[-] the average configuration of a polymer at thermal equilibrium is \emph{crumpled} (see Fig. \ref{fig:tessellated}); these are the configuration that give the dominant contribution to the entropy. Correspondingly, the average shape of the tessellated horizon is approximately \emph{spherical} (see Fig. \ref{fig:tessellated} for a picture of the shape reconstructed from the data $j_a \vec{n}_a$ of a random point in the ensemble); the spherical symmetry of the Schwarzschild macrostate is not imposed on the microstates, it is a property recovered in average at equilibrium;
	\item[-] in the LQG derivation of Black Hole entropy, the area law is reproduced and quantum corrections to it have been computed \cite{Corichi:2009wn}; such corrections are logarithmic in the area and feature a coefficient $-3/2$. The same $-3/2 \log$ corrections have been found in other approaches (in near-horizon conformal field theory calculations and in String theory); Carlip has pointed out that such coefficient may be universal \cite{Carlip:2000nv}.	
	
As already mentioned, the entropy of a polymer is proportional to its length; the proportionality coefficient $c$ depends on the monomer-chemistry. However, here too there are subleading corrections to the entropy; these corrections are logarithmic in the length $L$ and the proportionality coefficient here is known to be universal and can be computed. In particular, for a closed polymer chain the coefficient is $-3/2$,
\begin{equation}
S\;=\;c\,L\;-\frac{3}{2}\log L\;.
\label{eq:S pol}
\end{equation}
When translated back to the problem of Black Hole entropy, this relation immediately reproduces the logarithmic corrections found in LQG.
\end{itemize}
The LQG derivation of the entropy of a Schwarzschild Black Hole and of its leading-order quantum corrections is a rather complicated calculation \cite{Corichi:2009wn}. It is interesting to notice that, using semiclassical methods and the mapping to a polymer-physics problem, the results described above can be obtained via a very simple and familiar calculation. Such calculation is presented in next section.

\section{Computation of the entropy and of leading-order corrections}\label{sec:calculation}
The calculation of $W(A_H)$, the number of accessible configurations of the tessellated horizon, can be done in two steps: 
\begin{itemize}
	\item we first compute the number of microstates corresponding to an assignment of spins $j_a$ to the $N$ facets of the tessellated surface; this corresponds exactly to the problem of computing the number for conformations of a closed polymer chain with given monomer content;
	\item the second step is to take into account the fact that, within the ensemble of given average total area, the number of facets and the assignment of spins can vary. This is done computing the statistical distribution of spins that extremizes the entropy.
\end{itemize}
The two steps are discussed below, the first in section \ref{sec:calculation1} the second in section \ref{sec:calculation2}.

\subsection{Number of microstates for given spins/monomer lengths}\label{sec:calculation1}
Let us call $\Omega(j_1\mdots j_N)$ the number of microstates of a tessellated surface with $N$ facets with an assignment of spins $j_1\mdots j_N$. As explained in section \ref{sec:tessellated}, the tessellated surface arises from the semiclassical limit of a spin system. To each spin Hilbert space $\mc{H}^{(j_a)}$
we can associate a classical phase space with phase space volume $\frac{j_a+1/2}{2\pi}\int d\vec{n}_a$. The term in front of the integral guaranties that there is an integer number of Planck cells in phase space. The unconstrained integral over directions $\vec{n}_a$ gives the solid angle $4\pi$, and the phase space volume coincides with the dimension of a single spin Hilbert space, $\dim \mc{H}^{(j_a)}=2j_a+1$. Here we are interested in a system on $N$ spins that satisfy the condition (\ref{eq:q closure}). At the semiclassical level, this condition is the constraint (\ref{eq:j closure}) on the directions $\vec{n}_a$ and can be imposed as a delta-function in the $N$ integrals over the normals. It follows that $\Omega(j_1\mdots j_N)$ is given by
\begin{equation}
\Omega(j_1\mdots j_N)=\prod_{a=1}^N\Big(\frac{j_a+1/2}{2\pi}\Big)\; \int d\vec{n}_1\cdots d\vec{n}_N\;\;\delta\Big(\sum_{a=1}^N j_a \vec{n}_a\Big)\;.
\label{eq:Omega}
\end{equation}
This is also the formula for the number of conformations of a closed polymer chain with $N$ monomers of length $j_1\mdots j_N$. It provides a semiclassical Bohr-Sommerfeld approximation\footnote{There is an exact formula for the dimension of intertwiner space. It is given by a group integral of a product of $SU(2)$ characters, 
\begin{equation}
\dim \mc{H}_0^{(j_1\mdots j_N)}\;=
\;\int_{SU(2)}dg\;\prod_{a=1}^N\chi^{(j_a)}(g)\;=\;\frac{1}{\pi}\int_0^{2\pi}d\theta\; \sin^2\frac{\theta}{2}\;\prod_{a=1}^N\frac{\sin(2j_a+1)\theta/2}{\sin\theta/2}\;.
\label{eq:dimH}
\end{equation}
This is the formula generally used as starting point for the computation of Black Hole entropy in LQG.
} of the dimension of intertwiner space, $\dim \mc{H}_0^{(j_1\mdots j_N)}$.

There are two standard tricks that allow to compute the quantity (\ref{eq:Omega}) in the approximation we are interested in. The first is to write the delta function in Fourier transform. The second is to introduce \emph{occupation numbers}: we call $N_j$ the number of facets that have spin $j=\frac{1}{2},1,\frac{3}{2},\ldots$ Clearly, the occupation numbers sum up to the total number of facets, $\sum_j N_j = N$. We call $\Omega(N_j)$ the number of configurations for given occupation numbers. It is simply given by
\begin{equation}
\Omega(N_j)=\prod_j\Big(\frac{j+1/2}{2\pi}\Big)^{N_j}\; \int\frac{d\vec{p}}{(2\pi)^3}\;\prod_j \Big({\textstyle\int d\vec{n}\;}\, e^{i\vec{p}\cdot\vec{n}\,j}\Big)^{N_j}\;.
\label{eq:OmegaN}
\end{equation}
We are interested in a large classical Black Hole. Its horizon area is much larger that the Planck scale, $A_H\gg L_P^2$. In this situation, it is reasonable to assume that the average total number of facets $N$ of the tessellated horizon is large, and that the occupation numbers $N_j$ are either zero or large,
\begin{equation}
N_j\gg 1\;.
\label{eq:Ngg}
\end{equation}
Under this assumption, to be checked a posteriori, we can easily estimate the number of configurations $\Omega(N_j)$. The integral over $d\vec{n}$ is easily performed, the result is $4\pi \frac{\sin p j}{p j}$ where $p$ is the norm of the dummy variable $\vec{p}$. Now we can use the fact that, for large $N$, the following asymptotic formula holds
\begin{equation}
\Big(\frac{\sin x}{x}\Big)^N\;\approx \;e^{-\frac{1}{6}N x^2}\qquad \text{for}\quad N\gg 1 \;.
\label{eq:asymp}
\end{equation}
Therefore we are left with a $3d$ Gaussian integral over $\vec{p}$. This is easily done and the result is
\begin{equation}
\Omega(N_j)\approx\;\Big(\prod_{j=\frac{1}{2},1,\frac{3}{2},\ldots}(2j+1)^{N_j}\Big)\;\;\frac{1}{\big(\frac{2\pi}{3}\sum_j j^2 N_j\big)^{\frac{3}{2}}}\;.
\label{eq:O asym}
\end{equation}
The first term in parenthesis leads to the length-law for a polymer and to the area law for the Black Hole. The second term comes from the closure condition (\ref{eq:j closure}) imposed as a delta function in (\ref{eq:Omega}). Technically, it comes from the integration of the $3d$ Gaussian. This is how the exponent $-3/2$ arises. This is the $-3/2$ found in polymer-physics for closed chains, and in the LQG derivation of corrections to Black Hole entropy.

\subsection{Statistical distribution and occupation numbers}\label{sec:calculation2}
Let us call $W(N_j)$ the number of distinguishable microstates with occupation numbers $N_j$. To identify which states are distinguishable it is important to understand how the shape of the tessellated surface depends on the spins $j_a$ and the normals $\vec{n}_a$. This is discussed in detail in \cite{Bianchi:2010gc}. In particular, for a given ordered choice of normals, we can ask what happens if we exchange the spins associated to two faces. Clearly, if the value of the two spins is equal, the reconstructed shape is the same. On the other hand, exchanging two different spins, we obtain a new shape. Therefore, the number of distinguishable configurations for given occupation numbers is $\Omega(N_j)$ times the same factor that appears in Boltzmann statistics:
\begin{equation}
W(N_j)=\frac{\big(\sum_j N_j\big)!}{\prod_j N_j!}\;\Omega(N_j)\;.
\label{eq:WNj}
\end{equation}
We are interested in the distribution of the occupation numbers at equilibrium. The equilibrium distribution $N_j$ is the one that maximizes the entropy $S(N_j)=\kappa \log W(N_j)$ with the constraint that characterizes the ensemble. Here the constraint is that the average total area of the tessellated surface is fixed to be the horizon area $A_H$, Eq. (\ref{eq:<A>}). Such constraint can be introduced via a Lagrange multiplier $\mu$. It plays the role of chemical potential for the number of facets. The condition the occupation numbers have to satisfy is that the following variation with respect to $N_j$ vanishes\footnote{The standard procedure discussed here has been used in the LQG derivation of Black Hole entropy by Khriplovich \cite{Khriplovich:2001je} and by Ghosh and Mitra \cite{Ghosh:2006ph}. The novel feature here is that it is applied to $SU(2)$ boundary degrees of freedom, not to $U(1)$ as in the previous calculations.}:
\begin{equation}
0\;=\; \frac{\p}{\p N_j}\Big(S(N_j)\;+\;\mu\;\kappa\;\,\big(\;\frac{A_H}{8\pi\, \gamma L_P^2}-\sum_j \sqrt{j(j+1)}N_j\;\big)\Big)\;.
\label{eq:de S}
\end{equation}
We assume that the occupation numbers are large, $N_j\gg 1$, and use Stirling's approximation for the factorials. In the following we call $N_j^*$ the occupation numbers at equilibrium, i.e. when (\ref{eq:de S}) is satisfied, and $N^*=\sum_j N_j^*$ the average total number of facets. At leading order in large $N^*$, we have that Eq. (\ref{eq:de S}) reduces to the following expression for the probability distribution $p_j$
\begin{equation}
p_j\equiv\frac{N^*_j}{N^*}\approx (2j+1)\;e^{-\mu^*\,\sqrt{j(j+1)}}\;,
\label{eq:pj}
\end{equation} 
where the value $\mu^*$ of the Lagrange multiplier is determined by the requirement that the probability is normalized to unity, $\sum_j p_j\;=1$. This is exactly the same equation 
\begin{equation}
\textstyle 1=\sum_j\; (2j+1)\; e^{-\mu^*\,\sqrt{j(j+1)}}
\label{eq:Pell}
\end{equation}
that appears in the counting done via other methods in LQG. The solution can be found numerically, $\mu^*\simeq 1.722$.

The average total number of facets $N^*$ can be determined using Eq. (\ref{eq:<A>}). Defining the constant $\alpha^*=\sum_j \sqrt{j(j+1)}\;p_j\,\simeq 1.460$, we have that the average number of facets is approximately the area of the horizon in Planck scale units,
\begin{equation}
N^*\approx\frac{A_H}{8\pi\, \gamma L_P^2\;\alpha^*}\;.
\label{eq:Nstar}
\end{equation}
Therefore it is large for a classical Black Hole, as assumed above. Using Eqs. (\ref{eq:pj}) and (\ref{eq:Nstar}), we can determine the occupation numbers $N_j^*$ as a function of the horizon area,
\begin{equation}
N_j^*\approx \frac{(2j+1)\; e^{-\mu^*\,\sqrt{j(j+1)}}}{8\pi\, \gamma L_P^2\;\alpha^*}\;A_H\;.
\label{eq:Njstar}
\end{equation}
The entropy at equilibrium is given by $\kappa \log W(N_j^*)$, it a function of the horizon area, and it is straightforward to evaluate it using Eqs. (\ref{eq:WNj}) and (\ref{eq:Njstar}). It is given by
\begin{equation}
S=\frac{\kappa}{4\,L_P^2}\;\frac{\mu^*}{2\pi \gamma}\;A_H\;.
\label{eq:Sgamma}
\end{equation}
The area law (\ref{eq:S BH}) is reproduced. If we identify the Planck scale $L_P$ with $\sqrt{G \hbar}$ and require the proportionality constant to reproduce the $\kappa/(4\,G \hbar)$ appearing in Bekenstein-Hawking formula, we can determine the value of the Immirzi parameter to be $\gamma=\frac{\mu^*}{2\pi}\simeq 0.274$. This is exactly the same value of $\gamma$ determined in LQG (see \cite{Corichi:2009wn}, ENP approach).

Now some comments on the logarithmic corrections  to the entropy. They come from the next-to-leading contributions in the asymptotic expansion in large number of facets $N^*$. The probability distribution can be corrected iteratively in this expansion. Because of the $\log$ in the definition of the entropy, the only non-trivial contribution comes from the second terms on the right hand side of Eq. (\ref{eq:O asym}). Discarding constants that do not depend on the horizon area, the leading order correction is
\begin{equation}
\Delta S=-\frac{3}{2}\,\kappa \;\log\frac{A_H}{L_P^2}\;.
\label{eq:log}
\end{equation}
This is exactly the same correction found in the LQG literature by other methods, see \cite{Kaul:2000kf}.\\

To summarize, using semiclassical methods and the mapping to a polymer chain, we have easily derived the standard results about the entropy of a classical non-rotating Black Hole that are generally obtained in LQG via  more involved methods\footnote{The standard LQG derivation involves an exact counting of microstates. The approximation of large classical Black Hole is made only in the last stage of the calculation. This exact counting allows to study the fine-grained structure of the entropy for small Black-Holes where quantum effects are important. See \cite{Corichi:2009wn} for a review of these recent developments.}.\\

In next section I explore a more speculative direction. I consider a common phenomenon in polymer physics, take seriously the correspondence with the Black Hole, and use it to suggest a new way to describe the horizon of a rotating Black Hole in LQG.

\section{Polymer stretching and rotating Black Holes}\label{sec:stretching}
Suppose we stretch our polymer to an elongation $R$. Clearly, its entropy will decrease as for larger $R$ there are fewer microstates. For small elongation, we have that the number of microstates is Gaussianly distributed,
\begin{equation}
W(L,R)\approx \;W_0(L)\;\exp\big({-c\frac{R^2}{L}}\big)\;,
\label{eq:WLR}
\end{equation}
and the entropy depends on the stretching in a quadratic way,
\begin{equation}
S(L,R)=\;c\; \big(\,L\;-\;\frac{(2 R)^2}{4L}\big)\;.
\label{eq:SLR}
\end{equation}
This is in fact the origin of the elasticity of rubber \cite{DeGennes}. Moreover, for larger $R$, there is an \emph{extremal} value for the stretching: it has to be smaller than half the length of the polymer\footnote{Recall that we are considering a \emph{closed} chain.}. The question I want to discuss in this section is if there is an analogous phenomenon for Black Holes.\\

When using the mapping described in section \ref{sec:mapping}, the average tessellated surface corresponding to a stretched polymer chain has an ellipsoidal geometry. It turns out that the horizon of a Kerr Black Hole has this geometry when written in Kerr-Schild form (see \cite{Visser:2007fj}). 

Notice that the angular momentum $J$ of a rotating Black Hole is bounded by its extremal value,
\begin{equation}
GJ\leq (GM)^2\;,
\label{eq:Jextr}
\end{equation}
similarly to what happens to the elongation $R$ of the polymer.
Moreover, for given mass, the larger is the angular momentum of the Black Hole, the lower is its entropy. In particular, in the slowly rotating case, Smarr formula applies: the dependence of the entropy on the angular momentum is quadratic and given by
\begin{equation}
S(M,J)=\frac{\kappa}{G\hbar}4\pi\Big((GM)^2-\frac{(GJ)^2}{4(GM)^2}\Big)\;.
\label{eq:Smarr}
\end{equation}
When compared to Eq. (\ref{eq:SLR}), this expression of the entropy suggests the identification of the mass squared of the Black Hole with the length $L$ of the polymer, and the angular momentum $J$ with the elongation.

Now, let us go back to LQG and intertwiners. It turns out that we are led to a semiclassical description of a Kerr horizon that is related to an old idea proposed by Krasnov \cite{Krasnov:1998vc} (see also \cite{Sahlmann:2007jt}). A stretched closed polymer corresponds to an intertwiner with ``fixed intermediate recoupling spin''. The intermediate spin $J$ corresponds to the angular momentum of the Black Hole. The counting of microstates can be done as in section \ref{sec:calculation}: we divide the facets in two groups, associate occupation numbers $N_j$ and $N'_j$ to each group, and require that the normals $j_a\vec{n}_a$  sum up to $J \vec{e}_z$ for the first group, and to its opposite for the second. The calculation of the entropy at equilibrium is straightforward and the result is a quadratic dependence on the angular momentum $J$ as in Smarr formula (\ref{eq:Smarr}).

This direction looks promising, it would be interesting to explore it in more detail. In particular, developing it involves identifying a notion of mass associated to the horizon of the Black Hole, a notion that appears to be still missing in LQG.

\section{Conclusions}\label{sec:conclusions}
In this paper, I have discussed a semiclassical description of the horizon microstates counted in Loop Gravity. The description is in terms of shapes of a tessellated horizon. The problem of counting shapes is solved mapping it to an equivalent statistical mechanical problem: the counting of conformations of a closed polymer chain. Using familiar statistical mechanics methods, the Bekenstein-Hawking area law and the logarithmic corrections to it are easily derived. This provides a simpler semiclassical derivation of the LQG results on Black Hole entropy. The Immirzi parameter is found to have the same value $\gamma=0.274$ obtained by other methods. Moreover, the construction suggests a new way to describe in LQG the horizon of a rotating Black Hole and to determine its entropy.

It is interesting to compare the description presented here to other possibly related approaches. Long ago, in a pioneering work \cite{Bekenstein:1974jk}, Bekenstein proposed that the horizon surface is tessellated, with $N$ patches of area $L_P^2$ providing a Planck-scale cut-off. A definite degeneracy $k$ is assumed for each patch. From these assumptions, the area law can be straightforwardly derived: the number of microstates is $k^N$ and the entropy is $S=\frac{\log k}{L_P^2}A_H$. The picture that arises from this construction is close to the one described here: Loop Gravity provides a cut-off that is rather similar to this one; the difference is that, besides knowing about the area of the quantum patches, it contains information about the possible \emph{shapes} of the horizon via its extrinsic curvature.

Another possibly interesting relation is the one with the notion of quasinormal modes and their relevance in the derivation of Black Hole entropy \cite{York:1983zb}. Here the microstates are associated to the oscillation modes of the horizon shape. The description in terms of a tessellated horizon contains such modes. However, what is missing at the present stage is their dynamics. Ideas proper to the membrane paradigm \cite{Damour:1982mg} may be useful in this respect.

To conclude, some more speculative remarks. Recently Dreyer \cite{Dreyer:2002vy} has suggested that the Hod conjecture \cite{Hod:1998vk} about quasinormal modes and the area spectrum may be realized in Loop Gravity. For this to happen, the $\log 3$ appearing in the spectrum of quasinormal modes should arise from a spin $j=1$ dominance. It is interesting to notice that the method we have used in section \ref{sec:calculation} gives as by-product a derivation of the distribution of the occupation numbers at equilibrium. Therefore one can ask what is the dominating spin: from Eq. (\ref{eq:Njstar}), we find that in average $45\%$ of the facets have spins $j=\frac{1}{2}\,$, $26\%$ have spin $j=1\,$, $14\%$ have spin $j=\frac{3}{2}\,$, and $15\%$ have a larger spin $j>\frac{3}{2}$. The average spin is approximately one, $\langle j\rangle\simeq 1.05\,$.  There may be a relation here to be further explored.  

Finally, a rather natural question to ask is if there is a relation with the recent work of Verlinde on the origin of gravity \cite{Verlinde:2010hp}. There, the analogy with the polymer plays a central role. In particular one can ask if, in the setting discussed in this paper, there is an entropic force associated to the stretching of the polymer. Taking seriously the correspondence outlined in section \ref{sec:stretching}, one is lead to argue that the work needed to raise the angular momentum of a Black Hole (without changing its mass) has an entropic origin: it corresponds to a decrease in the number of accessible microstates.

\section*{Acknowledgments}
Special thanks to Carlo Rovelli, Alejandro Perez, Hal Haggard, Pietro Don\'a and Robert Littlejohn for many stimulating discussions. I also wish to thank Don Marolf, Bill Unruh, Niayesh Afshordi, Lee Smolin, Leonardo Modesto, Matteo Polettini and James D. Bjorken for useful comments and suggestions.
This work was supported by a Marie Curie IEF Fellowship within the 7th European Community Framework Programme.


\begin{thebibliography}{10}

\bibitem{Bekenstein:1973ur}
J.~D. Bekenstein, 
{\em Phys. Rev.} {\bf D7}
  (1973)
2333--2346.

S.~W. Hawking, 
{\em Commun. Math.
  Phys.} {\bf 43} (1975)
199--220.

J.~M. Bardeen, B.~Carter, and S.~W. Hawking,
 {\em Commun. Math. Phys.} {\bf 31} (1973)
161--170.

\bibitem{Wald:1999vt}
R.~M. Wald,
 {\em Living Rev. Rel.}
  {\bf 4} (2001) 6,
\href{http://arXiv.org/abs/gr-qc/9912119}{{\tt gr-qc/9912119}}.

\bibitem{Corichi:2009wn}
A.~Corichi, 
 {\em J. Phys. Conf. Ser.},{\bf D140}, 012006 (2008),
\href{http://arXiv.org/abs/0901.1302}{{\tt 0901.1302}}.

I.~Agullo, J.~Fernando~Barbero, E.~F. Borja, J.~Diaz-Polo, and E.~J.~S.  Villasenor, 
  \href{http://link.aps.org/doi/10.1103/PhysRevD.82.084029}{{\em Phys. Rev.} {\bf D82}} (2010)
084029.

\bibitem{Rovelli:1996dv}
C.~Rovelli, 
{\em Phys. Rev.
  Lett.} {\bf 77} (1996) 3288--3291,
\href{http://arXiv.org/abs/gr-qc/9603063}{{\tt gr-qc/9603063}}.

K.~V. Krasnov, 
{\em
  Phys. Rev.} {\bf D55} (1997) 3505--3513,
\href{http://arXiv.org/abs/gr-qc/9603025}{{\tt gr-qc/9603025}}.

L.~Smolin, 
{\em J. Math. Phys.} {\bf 36} (1995) 6417--6455,
\href{http://arXiv.org/abs/gr-qc/9505028}{{\tt gr-qc/9505028}}.

\bibitem{Ashtekar:1997yu}
A.~Ashtekar, J.~Baez, A.~Corichi, and K.~Krasnov, 
  {\em Phys. Rev. Lett.} {\bf 80} (1998) 904--907,
\href{http://arXiv.org/abs/gr-qc/9710007}{{\tt gr-qc/9710007}}.

A.~Ashtekar, J.~C. Baez, and K.~Krasnov, 
  {\em Adv. Theor. Math. Phys.} {\bf 4}
  (2000) 1--94,
\href{http://arXiv.org/abs/gr-qc/0005126}{{\tt gr-qc/0005126}}.

\bibitem{Engle:2009vc}
J.~Engle, A.~Perez, and K.~Noui, 
 {\em Phys. Rev. Lett.} {\bf 105} (2010) 031302,
\href{http://arXiv.org/abs/0905.3168}{{\tt 0905.3168}}.

J.~Engle, K.~Noui, A.~Perez, and D.~Pranzetti, 
 {\em Phys. Rev.}
  {\bf D82} (2010) 044050,
\href{http://arXiv.org/abs/1006.0634}{{\tt 1006.0634}}.

A.~Perez and D.~Pranzetti,
\href{http://arxiv.org/abs/arXiv:1011.2961}{{\tt 1011.2961}}.


\bibitem{Rovelli:1996ti}
C.~Rovelli, 
{\em Helv. Phys.
  Acta} {\bf 69} (1996) 582--611,
\href{http://arXiv.org/abs/gr-qc/9608032}{{\tt gr-qc/9608032}}.

C.~Rovelli, ``{Quantum gravity},'' Cambridge Univ. Pr. (2004).

\bibitem{Krasnov:2009pd}
K.~Krasnov and C.~Rovelli,
  {\em  Class. Quant. Grav.} {\bf 26} (2009) 245009,
\href{http://arXiv.org/abs/0905.4916}{{\tt 0905.4916}}.

\bibitem{Livine:2007vk}
E.~R. Livine and S.~Speziale, 
  {\em Phys. Rev.} {\bf D76} (2007) 084028,
\href{http://arXiv.org/abs/0705.0674}{{\tt 0705.0674}}.

F.~Conrady and L.~Freidel, 
  {\em J. Math. Phys.} {\bf 50} (2009) 123510,
\href{http://arXiv.org/abs/0902.0351}{{\tt 0902.0351}}.

L.~Freidel, K.~Krasnov, and E.~R. Livine, 
  {\em Commun. Math. Phys.} {\bf 297} (2010) 45--93,
\href{http://arXiv.org/abs/0905.3627}{{\tt 0905.3627}}.

L.~Freidel and E.~R. Livine, 
{\em J. Math. Phys.} {\bf 51} (2010) 082502,
\href{http://arXiv.org/abs/0911.3553}{{\tt 0911.3553}}.

\bibitem{Bianchi:2010gc}
E.~Bianchi, P.~Don\'a, and S.~Speziale, 
\href{http://arXiv.org/abs/1009.3402}{{\tt 1009.3402}}.

\bibitem{Hal}
E.~Bianchi and H.~Haggard, 
{\em to appear (2010)}.

\bibitem{Sorkin:1983a}
R.~D. Sorkin, 
{\em In Bertotti et al. (Eds.), 10th Int.
  Conf. on Gravitation (Padova, 1983), p.734}

R.~D. Sorkin, 
{\em Stud. Hist. Philos.
  Mod. Phys.} {\bf 36} (2005) 291--301,
\href{http://arXiv.org/abs/hep-th/0504037}{{\tt hep-th/0504037}}.

\bibitem{York:1983zb}
J.~W. York, Jr., 
\href{http://link.aps.org/doi/10.1103/PhysRevD.28.2929}{{\em Phys.
  Rev.} {\bf D28} (1983)
2929.}

\bibitem{Jacobson:2005kr}
T.~Jacobson, D.~Marolf, and C.~Rovelli, 
   {\em Int. J. Theor. Phys.} {\bf 44} (2005) 1807--1837,
\href{http://arXiv.org/abs/hep-th/0501103}{{\tt hep-th/0501103}}.

\bibitem{Damour:1982mg}
T.~Damour, 
{\em \href{http://www.ihes.fr/~damour/Articles/surfaceeffects.pdf}{Proceedings} of the
  Second Marcel Grossmann Meeting, (ed. R. Ruffini,
  North Holland, 1982) pp 587- 608;
  } .

K.~S. Thorne, R.~H. Price, and D.~A. Macdonald, (Eds.),
  ``{Black Holes: the membrane paradigm},''
   Yale Univ. Pr. (1986).

M.~Parikh and F.~Wilczek, 
{\em Phys.
  Rev.} {\bf D58} (1998) 064011,
\href{http://arXiv.org/abs/gr-qc/9712077}{{\tt gr-qc/9712077}}.

\bibitem{Maggiore:1994ww}
M.~Maggiore, 
 {\em Nucl. Phys.} {\bf
  B429} (1994) 205--228,
\href{http://arXiv.org/abs/gr-qc/9401027}{{\tt gr-qc/9401027}}.

A.~Buonanno, M.~Gattobigio, M.~Maggiore, L.~Pilo, and C.~Ungarelli,
  {\em Nucl. Phys.} {\bf
  B451} (1995) 677--698,
\href{http://arXiv.org/abs/gr-qc/9504020}{{\tt gr-qc/9504020}}.

\bibitem{Rovelli:1994ge}
C.~Rovelli and L.~Smolin, 
{\em Nucl. Phys.} {\bf B442} (1995) 593--622,
\href{http://arXiv.org/abs/gr-qc/9411005}{{\tt gr-qc/9411005}}.

A.~Ashtekar and J.~Lewandowski, 
 {\em Class. Quant. Grav.} {\bf 14} (1997) A55--A82,
\href{http://arXiv.org/abs/gr-qc/9602046}{{\tt gr-qc/9602046}}.

\bibitem{Minkowski}
H.~Minkowski, 
 {\em
  Nachr. Ges. Wiss., G\"ottingen, 1897, 198-219}.

\bibitem{Major:1999mc}
S.~A. Major, 
{\em Class. Quant.
  Grav.} {\bf 16} (1999) 3859--3877,
\href{http://arXiv.org/abs/gr-qc/9905019}{{\tt gr-qc/9905019}}.

\bibitem{Kapovich}
M.~Kapovich and J.~J. Millson, 
{\em J. Differential Geom. {\bf 44}, 3 (1996), 479-513.}

\bibitem{Charles}
L.~Charles, 
{\em Asian J. Math.} 14-1 (2010), 109-152
 \href{http://arxiv.org/abs/0806.1585}{\tt [arXiv:0806.1585]}.

\bibitem{DeGennes}
P.~G. De~Gennes, 
``Scaling Concepts in Polymer Physics,''
 {\em Cornell
  Univ. Pr.} (1979).

\bibitem{Carlip:2000nv}
S.~Carlip, 
   {\em Class. Quant. Grav.} {\bf 17} (2000) 4175--4186,
\href{http://arXiv.org/abs/gr-qc/0005017}{{\tt gr-qc/0005017}}.

\bibitem{Khriplovich:2001je}
I.~B. Khriplovich, 
  {\em Phys. Lett.} {\bf B537} (2002) 125--129,
\href{http://arXiv.org/abs/gr-qc/0109092}{{\tt gr-qc/0109092}}.

\bibitem{Ghosh:2006ph}
A.~Ghosh and P.~Mitra,
   {\em Phys. Rev.} {\bf D74} (2006) 064026,
\href{http://arXiv.org/abs/hep-th/0605125}{{\tt hep-th/0605125}}.

\bibitem{Kaul:2000kf}
R.~K. Kaul and P.~Majumdar, 
  {\em Phys. Rev. Lett.} {\bf 84} (2000) 5255--5257,
\href{http://arXiv.org/abs/gr-qc/0002040}{{\tt gr-qc/0002040}}.

I.~Agullo, G.~J. Fernando~Barbero, E.~F. Borja, J.~Diaz-Polo, and E.~J.~S.
  Villasenor,
  {\em Phys. Rev.} {\bf D80} (2009) 084006,
\href{http://arXiv.org/abs/0906.4529}{{\tt 0906.4529}}.

\bibitem{Visser:2007fj}
M.~Visser, 
\href{http://arXiv.org/abs/0706.0622}{{\tt 0706.0622}}, in D.~L.~Wiltshire,
  M.~Visser, S.~M.~Scott, 
  Cambridge Univ. Pr. (2009).

\bibitem{Krasnov:1998vc}
K.~Krasnov, 
{\em Class.
  Quant. Grav.} {\bf 16} (1999) L15--L18,
\href{http://arXiv.org/abs/gr-qc/9902015}{{\tt gr-qc/9902015}}.

\bibitem{Sahlmann:2007jt}
H.~Sahlmann, 
{\em Class. Quant.
  Grav.} {\bf 25} (2008) 055004,
\href{http://arXiv.org/abs/0709.0076}{{\tt 0709.0076}}.

\bibitem{Bekenstein:1974jk}
J.~D. Bekenstein, 
{\em
  Lett. Nuovo Cim.} {\bf 11} (1974)
467.

J.~D. Bekenstein, 
proceeedings BSCG'98,
\href{http://arXiv.org/abs/gr-qc/9808028}{{\tt gr-qc/9808028}}.

\bibitem{Dreyer:2002vy}
O.~Dreyer, 
  {\em Phys. Rev. Lett.} {\bf 90} (2003) 081301,
\href{http://arXiv.org/abs/gr-qc/0211076}{{\tt gr-qc/0211076}}.

\bibitem{Hod:1998vk}
S.~Hod, 
  {\em Phys. Rev. Lett.} {\bf 81} (1998) 4293,
\href{http://arXiv.org/abs/gr-qc/9812002}{{\tt gr-qc/9812002}}.

\bibitem{Verlinde:2010hp}
E.~P. Verlinde, 
  \href{http://arXiv.org/abs/1001.0785}{{\tt 1001.0785}}.

\end{thebibliography}

\providecommand{\href}[2]{#2}\begingroup\raggedright\endgroup

\end{document}